\documentclass[aip,jcp,reprint]{revtex4-1}
\usepackage{epsfig}
\usepackage{color}
\usepackage{amsmath}
\usepackage{amsfonts}
\usepackage{natbib}
\usepackage{notes2bib}

\newcommand{\rb}{\mathbf{r}}

\newcommand{\avg}[1]{\left<#1\right>}
\newcommand{\len}[1]{\left|#1\right|}
\newcommand{\brac}[1]{\left[#1\right]}

\newcommand{\para}[1]{\left(#1\right)}

\DeclareMathOperator{\Tr}{Tr}

\newcommand{\ie}{\emph{i.e.}}

\newcommand{\etal}{\emph{et al.}}

\newcommand{\rhoq}{\ensuremath{\rho^q}}

\newcommand{\rhob}{\rho_{\rm B}}

\newcommand{\kT}{\ensuremath{k_{\rm B}T}}

\newcommand{\Qb}{\ensuremath{\mathcal{Q}}}

\newcommand{\Rbar}{\overline{\textbf{R}}}

\begin{document}


\title{Distributed Charge Models of Liquid Methane and Ethane for Dielectric Effects and Solvation}



\author{Atul C. Thakur}
\author{Richard C. Remsing}
\email[]{rick.remsing@rutgers.edu}
\affiliation{Department of Chemistry and Chemical Biology, Rutgers University, Piscataway, NJ 08854}




\begin{abstract}
Liquid hydrocarbons are often modeled with fixed, symmetric, atom-centered
charge distributions and Lennard-Jones interaction potentials that reproduce many properties
of the bulk liquid.
While useful for a wide variety of applications, such models cannot capture dielectric effects
important in solvation, self-assembly, and reactivity. 
The dielectric constants of hydrocarbons, such as methane and ethane, physically arise from electronic
polarization fluctuations induced by the fluctuating liquid environment. 
In this work, we present non-polarizable, fixed-charge models of methane and ethane that break the charge symmetry
of the molecule to create fixed molecular dipoles, the fluctuations of which reproduce the experimental dielectric constant.
These models can be considered a mean-field-like approximation that can be used to include dielectric effects
in large-scale molecular simulations of polar and charged molecules in liquid methane and ethane. 
We further demonstrate that solvation of model ionic solutes and a water molecule in these fixed-dipole
models improve upon dipole-free models.
\end{abstract}


\maketitle

\raggedbottom

\section{Introduction}

Understanding the liquid-state properties of hydrocarbons
is important for applications in the petrochemical industry~\cite{sattler2014catalytic,FARAMAWY201634,WOOD2012196},
their use as solvents for synthesis and separations, and
as general models for simple, non-associating liquids~\cite{chandler1987introduction,TheorySimpLiqs}.
Interest in the simplest of these liquids
has been reinvigorated by the discovery of
methane/ethane lakes on the cold ($\sim94$ K) surface of the Saturnian moon Titan~\cite{TitanLakesReview,NIXON201850,life6010008,Cordier_2013,NatureAstronomy:TitanLakes,TitanLakeComp,Horst:GeophysRes,Tokano_2006}. 
The existence of liquid reservoirs on Titan's surface, combined with its
rich atmospheric chemistry, has led many to hypothesize that the hydrocarbon lakes could harbor
prebiotic chemistry and even non-aqueous life~\cite{Sagan_Titan,TitanChapter,Neish_2018,Raulin_2012,KAWAI201920}. 
However, any such chemistry would be vastly different than similar processes in aqueous environments,
and a fundamental, molecular-scale understanding is necessary,
beginning with characterizing solvation in methane and ethane~\cite{MCKAY1996741,kawai2013titan,TitanChapter,Acetylene_Solubility,Cornet:GeophysRes,AcetonitrilePMF}. 
Such a microscopic picture of cryogenic hydrocarbon solutions can be provided by
molecular simulations, but there remains a need to make these simulations
efficient and predictive. 
One difficulty presented by modeling liquid hydrocarbons is a description of their dielectric properties. 
United-atom models combine the carbon and hydrogen atoms into single sites with intermolecular
interactions described by Lennard-Jones (LJ) potentials and cannot describe dielectric effects by construction~\cite{TrappeUA}. 
Most atomically-detailed molecular models of hydrocarbons describe the intermolecular interactions through
atom-based LJ and electrostatic interactions, with the latter achieved by
assigning a fixed set of point-charges to each molecule~\cite{righini1981intermolecular,OPLS,AMBERFF-2003}.
These and similar models have been reasonably successful, and can adequately describe many aspects of
the structure, dynamics, and thermodynamics of liquid and solid hydrocarbons,
including at conditions similar to those on Titan~\cite{righini1981intermolecular,FIRANESCU2011779,Luckhaus:MolPhys,bounds1980molecular,chen2002simulating}. 
However, symmetric fixed-charge and charge-free models of methane and ethane cannot readily describe
dielectric effects. 
Methane and ethane do not have a permanent dipole moment, due to symmetry,
and consequently any symmetric and rigid fixed charge model yields a dielectric constant of unity.
Therefore, these standard models cannot properly describe the response of hydrocarbon solvents
to polar and charged solutes. 
Physically, the dielectric responses of methane and ethane arise from
their polarizabilities.
The relevant dipole fluctuations can be accounted for by polarizable and ab initio models~\cite{vorobyov2005polarizable,davis2008revised,mcgrath2011vapor,richters2013liquid,Lemkul:2016aa,pruteanu2020squeezing}.
However, polarizable models can be difficult to parameterize and
are more expensive than the fixed charge models discussed above. 
An intermediate class of models with fixed charges and the ability to
describe dielectric effects was introduced by Fennell~\etal,
referred to as dielectric corrected (DC) models~\cite{dillmodels}. 
For symmetric molecules without a permanent dipole moment,
a DC model breaks the molecular charge symmetry to create a fixed dipole
moment, which is parameterized to reproduce the dielectric constant of the liquid phase. 
%

\begin{figure*}[tb]
\begin{center}
\includegraphics[width=0.7\textwidth]{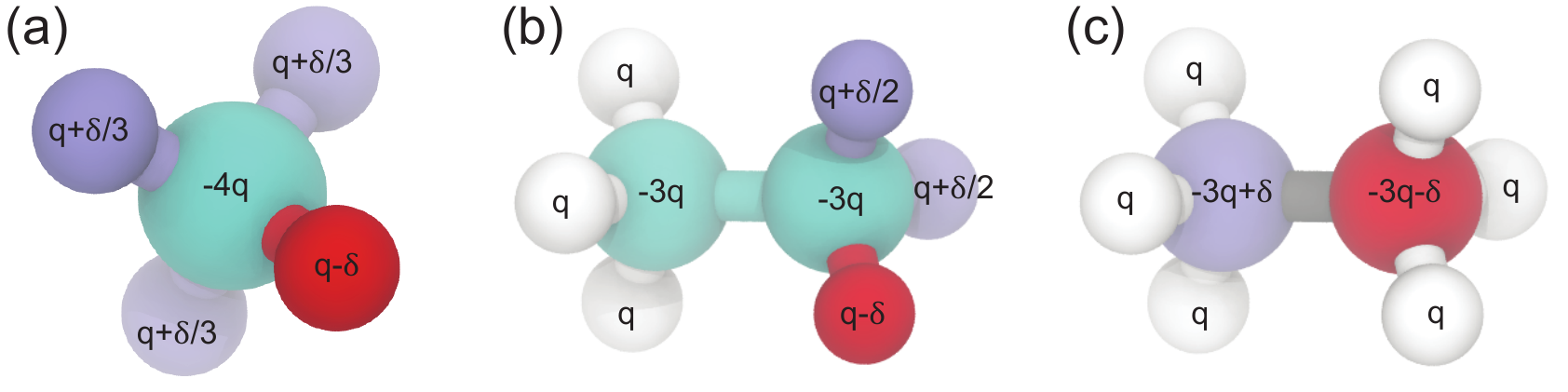}
\end{center}
\caption
{
Schematics of the charge distributions for (a) the DC model of methane and the (b) DC and (c) DC2 models of ethane.
}
\label{fig:models}
\end{figure*}

%
In this work, we present DC models for liquid methane and ethane at
Titan surface conditions.
In addition to describing dielectric constants of the pure liquids,
we find that the DC models provide a good description of the
dielectric constant of methane/ethane mixtures. 
We also demonstrate that these models yield structure and dynamics
in good agreement with the original, dipole-free models, such that the DC
models provide a reasonable description of the two bulk liquids. 
We then turn to the solvation of model solutes.
We first investigate hard sphere solvation and the corresponding liquid
density fluctuations, demonstrating that all models studied here
provide good descriptions of apolar solvation.
Then, we investigate the solvation of charged hard spheres as model ionic solutes
and the solvation of a water molecule, a polar solute.
In this case, the DC models provide very different results than
the symmetric, dipole-free models, because the DC models
exhibit a larger dielectric response. 
Our results suggest that DC models can be used in place of traditional dipole-free models
to predict solvation thermodynamics of polar and charged species in
hydrocarbon solutions, like those on the surface of Titan.

\section{Simulation Details}
All simulations were performed with GROMACS 2020~\cite{gmx4ref,Berendsen1995,Spoel2005}.
Simulations of pure liquid methane and ethane were performed with 697 molecules,
and mixture simulations were performed with 697 molecules of one {type} and 299 molecules of the other.
After constructing the simulation cells and performing an energy minimization,
the systems were equilibrated for 1~ns in the canonical ensemble, followed by
equilibration of at least 10~ns in the isothermal-isobaric (NPT) ensemble.
Statistics were gathered over production runs of at least 50~ns in length in the NPT ensemble. 
{Equations of motion were integrated using the leapfrog algorithm with a 1~fs timestep.}
A constant temperature of 94~K was maintained using a Nos\'{e}-Hoover thermostat~\cite{Nose1984,Hoover1985}
and a constant pressure of 1~bar was maintained using an Andersen-Parrinello-Rahman barostat~\cite{AndersenBaro,Parrinello-Rahman}. 
Short-range interactions (LJ and Coulomb) were truncated at 1~nm, with {standard long-range tail corrections} applied for the LJ contribution
to the energy and pressure. 
Long-range electrostatic interactions were evaluated using the particle mesh Ewald method~\cite{PME}.
All C-H bond lengths were constrained using the LINCS algorthim~\cite{PLINCS}.
All bond, angle, and LJ parameters were taken from the OPLS force field~\cite{OPLS}, to which we compare the results of the DC models.

In order to simulate a hard-sphere-like solute with a radius of 3~\AA \ in methane and ethane,
we created a non-interacting dummy particle, fixed at the center of the box,
and we biased the coordination number of this particle with a harmonic potential using PLUMED~\cite{PLUMED}. 
For the harmonic potential $U(\tilde{N})=\kappa/2(\tilde{N}-\tilde{N}^*)^2$, where $\tilde{N}$
is a smoothed variant of the coordination number necessary for biasing~\cite{INDUS,INDUS-2},
we set $\kappa=5$~kJ/mol and $\tilde{N}^*=-20$ in order to exclude all solvent molecules from
within 3~\AA \ of the solute particle. 
The biasing potential was applied to solvent carbon atoms only. 
Simulations of the methane liquid-vapor interface were performed in the canonical ensemble
using a Nos\'{e}-Hoover thermostat~\cite{Nose1984,Hoover1985}.
A liquid slab was created by elongating the $z$-axis of an equilibrated bulk simulation by a factor of three.
{Short-range interactions were truncated at 1~nm, such that the effects of the LJ tails on the interface are neglected beyond that length.}
Long-range electrostatic interactions were evaluated using the particle mesh Ewald method~\cite{PME} in conjunction
with the correction of Yeh and Berkowitz for slab-like systems~\cite{Ewald3DC},
and all other simulation parameters followed those of the bulk systems. 
{To simulate solvation of a water molecule in liquid methane,
we employed the widely-used extended simple point charge (SPC/E) model of water~\cite{SPCE}.
This model includes a LJ potential centered on the oxygen site and point charges $q_{\rm H}$ and $q_{\rm O}=-2q_{\rm H}$ located on
the hydrogen and oxygen sites, respectively.
The LJ interactions between methane sites and the SPC/E water molecule were determined using standard OPLS combining rules.
All simulation parameters followed those for the bulk and hard sphere systems detailed above. 
Simulations of charging the SPC/E models from zero (LJ solute only) to the fully charged water model were equilibrated for a minimum of 5~ns
and production runs were 10~ns in length.}
%

\section{Static Dipolar Charge Distributions can Reproduce the Dielectric Constant}

Due to symmetry, both methane and ethane do not have static molecular dipole moments,
so that the dielectric constant is determined by electronic polarization fluctuations. 
Here, we develop models with fixed, effective dipole moments --- using point charges distributed over the molecular sites ---
that can reproduce the experimental dielectric constant of each liquid. 
This approach can be considered a mean-field-like approximation to the polarization fluctuations
akin to the distributed-dipole DC models of Fennell~\etal~\cite{dillmodels}.
{However, we do not attempt to fit the temperature dependence of the dielectric constant, as done by Fennell~\etal~\cite{dillmodels},
which required altering LJ parameters in addition to atomic charges. 
Instead, we fit the dielectric constant at a single state point and minimally perturb the model by making small changes in the atomic charges only.}

We tune the fixed point charges on atomic sites according to the schemes in Fig.~\ref{fig:models},
where dipoles are created using {a shift parameter $\delta=q$ that modifies the charge on specified sites.}
Ethane presents more freedom in the choice of charge distribution, and so we parameterize two models: DC and DC2.
The DC ethane model has a charge distribution similar to the DC methane model, while the DC2 model
creates a permanent dipole moment using the carbon atoms only. 
{We note that these choices are not unique, and equivalent results can be obtained using other
charge distributions with roughly the same dipole moment. For example, we also parameterized
a methane model with $q=0.06$ and $\delta=0.0441$ that yields properties equivalent to the DC model
at the focus of this work.}

{The magnitude of $q$ is optimized to match the experimental dielectric constants at 94~K.
These models may not be readily transferable to different state points, because only the single state of interest
was considered when determining $q$, as mentioned above. 
The resulting parameters are listed in Table~\ref{tab:molec}.
The resulting charges and dipole moments are smaller than those determined by Fennell~\etal \ for CCl$_4$, for example~\cite{dillmodels},
and the dipole is similar to that of the Fox and Kollman model for CCl$_4$~\cite{fox1998application}.
Altering the atomic charges to create a dipole moment also changes the quadrupole tensor of the molecule, where $\mathbf{Q}$ and $\mathcal{Q}$
are respectively the primitive and traceless quadrupole tensors. 
Therefore, we list the trace of $\mathbf{Q}$, which is used to estimate the Bethe potential discussed below in the context of ion solvation~\cite{Remsing:JPCL:2014},
and the off-diagonal elements of $\mathcal{Q}$, indicated by $\tilde{\mathcal{Q}}$. 
}

\begin{figure}[tb]
\begin{center}
\includegraphics[width=0.4\textwidth]{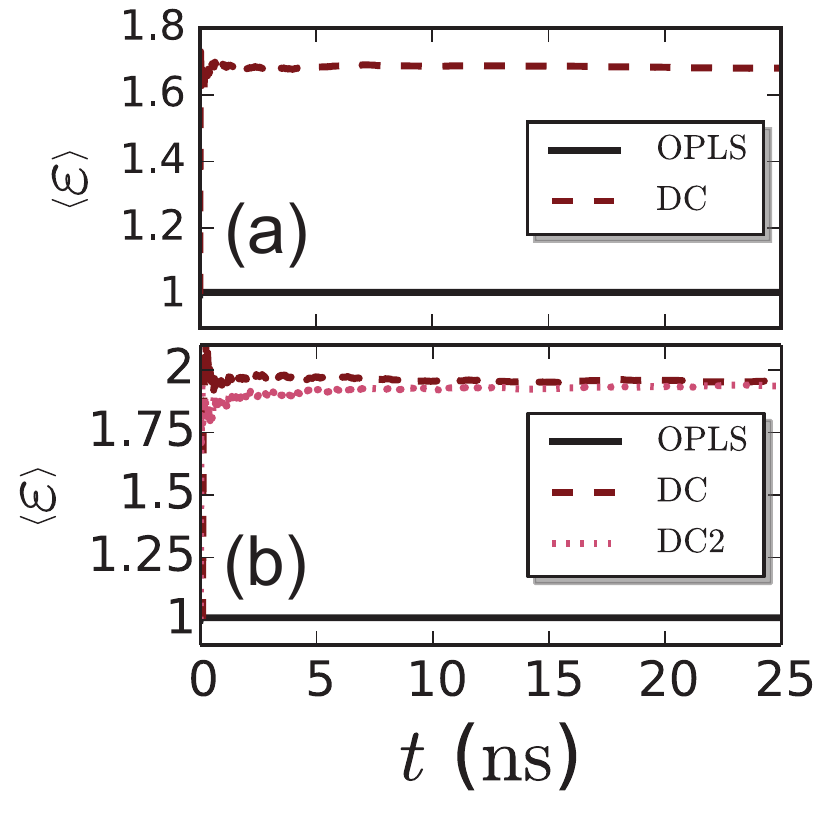}
\end{center}
\caption
{
Running averages of the dielectric constant in the (a) methane and (b) ethane models studied here,
shown for the first 25~ns of a 50~ns trajectory.
}
\label{fig:runavg}
\end{figure}

The dielectric constants
and bulk densities of those models are listed in Table~\ref{tab:bulk}, where the dielectric constants were determined according to
\begin{equation}
\varepsilon = 1+\frac{4\pi\beta}{3\avg{V}}\avg{(\delta \mathbf{M})^2},
\end{equation}
where $\beta^{-1}=\kT$ is the product of Boltzmann's constant and the temperature, $\avg{\cdots}$ indicates an ensemble average,
$V$ is the volume of the simulation cell, $\delta \mathbf{M} = \mathbf{M} - \avg{\mathbf{M}}$, and $\mathbf{M}$ is the total dipole moment of the system. 
The running average of $\varepsilon$ is shown in Fig.~\ref{fig:runavg} for all models studied. 
The dielectric constants of the DC models are in good agreement with those determined experimentally.
The OPLS models have dielectric constants close to unity, with deviations coming from intramolecular H-C-H and H-C-C angle fluctuations. 
The bulk densities (Table~\ref{tab:bulk}) show that the density is only slightly increased
in the DC models, in comparison to the OPLS models, in agreement with previous work
that showed that reasonable atomic charges have little impact on the thermodynamic properties of liquid alkanes~\cite{kaminski1994free,Chen:1999aa}.
%
%

\begin{table}
\caption{
\label{tab:molec}
Molecular properties for the models studied here: charge, $q$ ($e_0$), shift parameter, $\delta$ ($e_0$), 
dipole moment, $\mu$ (D), trace of the primitive quadrupole tensor, $\Tr{\mathbf{Q}}$ (D$\cdot$\AA),
and the off-diagonal element of the traceless quadrupole tensor, $\tilde{\Qb}$ (D$\cdot$\AA).
}
\vspace{0.6cm}
\centering
\begin{tabular*}{\hsize}{@{\extracolsep{\fill}}lcccccc}
\hline
Liquid & Model & $q$ & $\delta$ & $\mu$ & $\Tr{\mathbf{Q}}$ & $\tilde{\Qb}$ \\
\hline
Methane & OPLS & 0.06 & 0.0 & 0.0 &  1.370 & 0.0 \\[1.0ex]
& DC & 0.0444 & 0.0444  & 0.31 & 1.014 & -0.338 \\[2.0ex]
Ethane & OPLS & 0.06 & 0.0 & 0.0 &  3.00 & 0.264 \\[1.0ex]
& DC & 0.0576 & 0.0576 & 0.426 & 2.88 & 1.04 ($xy$/$xz$), -1.31 ($yz$) \\[1.0ex]
& DC2 & 0.06 & 0.06 & 0.435 & 2.35 & -0.393 \\[1.0ex]
\end{tabular*}
\end{table}

\begin{table}
\caption{
\label{tab:bulk}
Bulk properties for the models studied here: 
predicted dielectric constants and densities~(kg/m$^3$) for the methane models studied here.
Experimental dielectric constants~\cite{amey1964dielectric,EthaneDielectric,DielectricConstantsMixtures} and densities~\cite{NISTChemistryWebBook} are also listed.
{Error estimates are listed in parentheses and correspond to the standard deviation
among three independent simulations.}
}
\vspace{0.6cm}
\centering
\begin{tabular*}{\hsize}{@{\extracolsep{\fill}}lccc}
\hline
Liquid & Model & $\varepsilon$ & $\rhob$  \\
\hline
Methane & OPLS & 1.0073 (0.0007) & 499.2 (0.6) \\[1.0ex]
& DC &  1.680 (0.002) & 501.36 (0.03) \\[1.0ex]
& Exp. & 1.67 & 447.04 \\[2.0ex]
Ethane & OPLS & 1.0090 (0.0001) &  668.38 (0.06) \\[1.0ex]
& DC & 1.95 (0.01) & 664.6 (0.2) \\[1.0ex]
& DC2 & 1.94  (0.01) & 663.46 (0.08)  \\[1.0ex]
& Exp. & 1.94 & 647.65
\end{tabular*}
\end{table}

\begin{figure}[tb]
\begin{center}
\includegraphics[width=0.45\textwidth]{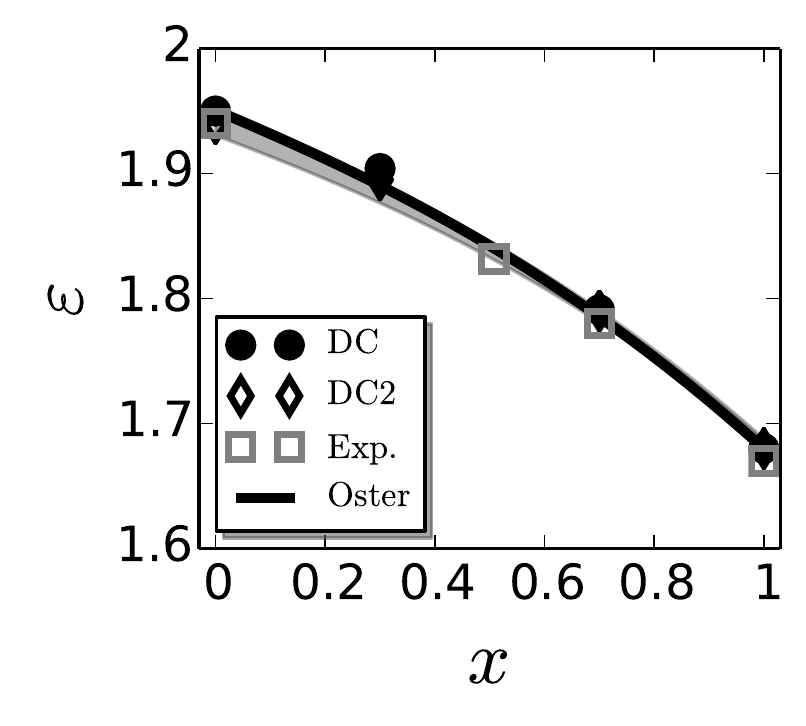}
\end{center}
\caption
{
Dielectric constant of methane-ethane mixtures as a function of the methane mole fraction, $x$,
determined via simulation with the DC models developed here and determined by experiments~\cite{DielectricConstantsMixtures}.
Also shown are the predictions from Oster's formula~\cite{Oster}, Eq.~\ref{eq:oster},
with the shaded region indicating the range of predictions consistent with the error bars. 
}
\label{fig:mixeps}
\end{figure}

%
Although the DC models were parameterized to match the dielectric constant of
pure liquid methane and ethane, they also make reasonable predictions for
the dielectric constant of their mixtures. 
To demonstrate this, we performed simulations of methane-ethane mixtures
with methane mole fractions of $x=0.3$ and $x=0.7$. 
The dielectric constants as a function of $x$ are shown in Fig.~\ref{fig:mixeps}, along with
available experimental data points. 
We also show the predictions of Oster's formula for the dielectric constant of mixtures~\cite{Oster},
\begin{equation}
\frac{\varepsilon(x)-1}{\varepsilon(x)+2} =  \sum_i x_i \frac{\rhob(x)}{\rho_{\rm B,i}} \frac{\varepsilon_i-1}{\varepsilon_i+2},
\label{eq:oster}
\end{equation}
where $x_i$ is the mole fraction of component $i$,
$\rhob(x)$ is the number density of the mixture $x$,
$\rho_{\rm B,i}$ is the bulk density of pure component $i$,
and $\varepsilon_i$ is the dielectric constant of pure component $i$.
To determine $\varepsilon$ for intermediate mole fractions, we fit the density to a quadratic function of $x$
and use this as input to Eq.~\ref{eq:oster}.
%

%
The concentration-dependence of the dielectric constant, shown in Fig.~\ref{fig:mixeps}, is in good agreement with experimental results
and the predictions of Eq.~\ref{eq:oster}.
The Oster equation is anticipated to be accurate for methane-ethane mixtures, because it is
an extension of the Clausius-Mossotti formula~\cite{Zangwill},
which has been shown to be accurate for pure methane and ethane liquids~\cite{EthaneDielectric,DielectricConstantsMixtures}.
The good agreement among the predictions of the DC models, Eq.~\ref{eq:oster},
and experiments suggests that these models can be accurately used to simulate
dielectric effects at a range of concentrations, including the ranges anticipated for Titan's lakes. 
%

\section{Liquid-State Structure and Dynamics} 
The OPLS models of methane and ethane yield accurate predictions for the
structure and dynamics of these liquids.
In this section, we demonstrate that creating the DC models of methane and ethane
leaves the structure and dynamics nearly unchanged. 
%

\begin{figure}[tb]
\begin{center}
\includegraphics[width=0.48\textwidth]{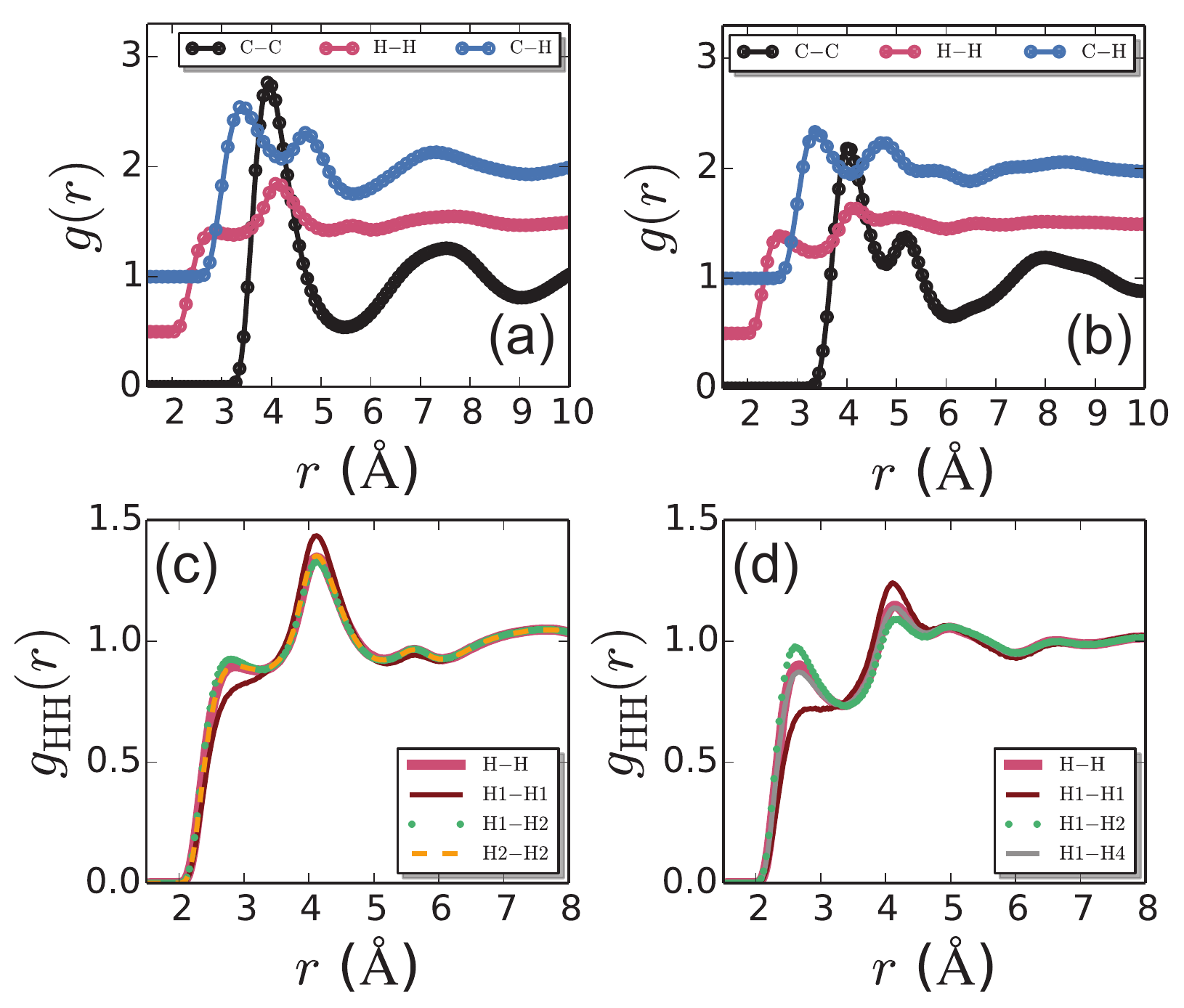}
\end{center}
\caption
{
Radial distribution functions, $g(r)$, for C-C, H-H, and C-H (intermolecular) correlations
in liquid (a) methane and (b) ethane.
Lines indicate $g(r)$ obtained using the dipole-free, OPLS model, and those
for the DC models are shown with data points. 
The H-H and C-H results are shifted vertically by 0.5 and 1, respectively. 
{Also shown are select $g(r)$ for correlations between H sites in (c) methane and (d) ethane.
H-H indicates to the OPLS and site-averaged DC $g(r)$. 
For DC methane, (c), H1 is the site with zero charge and H2 indicates the other H sites. 
For DC ethane, (d), H1 is the site with zero charge, H2 indicates the sites with charge $q+\delta/2$,
and H4 indicates the H sites with charge $q$ that are bonded to the other carbon atom in the molecule.}
}
\label{fig:gr}
\end{figure}

%
We characterize the structure of liquid methane and liquid ethane through site-site
pair distribution functions, $g_{\alpha \gamma}(r)$, where $\alpha$ and $\gamma$ represent
atomic sites. 
The carbon-carbon (CC), hydrogen-hydrogen (HH), and carbon-hydrogen (CH) pair distribution functions
of liquid methane and ethane are shown in Fig.~\ref{fig:gr} for the OPLS and DC models. 
The various $g_{\alpha\gamma}(r)$ are essentially identical for the two models. 
This illustrates that the small change in charge distributions necessary to obtain the experimental
dielectric constant does not significantly change the structure of the bulk liquid, resulting
in fixed-charge models with accurate structure and dielectric properties. 
The DC2 model yields $g_{\alpha\gamma}(r)$ indistinguishable from the OPLS and DC models and are not shown for clarity. 
{Although the pair distribution functions averaged over all sites are equivalent in the OPLS and DC models,
those between nonequivalent sites of the DC models can differ. 
For example, $g_{\rm HH}(r)$ differs for the two types of H sites in DC methane, Fig.~\ref{fig:gr}c. 
The correlations between like charged sites are diminished, while
correlations between nonequivalent sites are slightly increased beyond the average. 
This is indicative of dipolar correlations expected of a dielectric fluid. 
Similar correlations between nonequivalent H sites are also found in DC ethane, as highlighted by select $g_{\rm HH}(r)$ in Fig.~\ref{fig:gr}d. 
Correlations between equivalent sites, e.g. H1-H1, are diminished with respect to the average, H-H,
while significant correlations between nonequivalent sites can be enhanced beyond the site-average, H1-H2,
again indicative of dipolar correlations typical of dielectric media. 
}

\begin{figure}[tb]
\begin{center}
\includegraphics[width=0.48\textwidth]{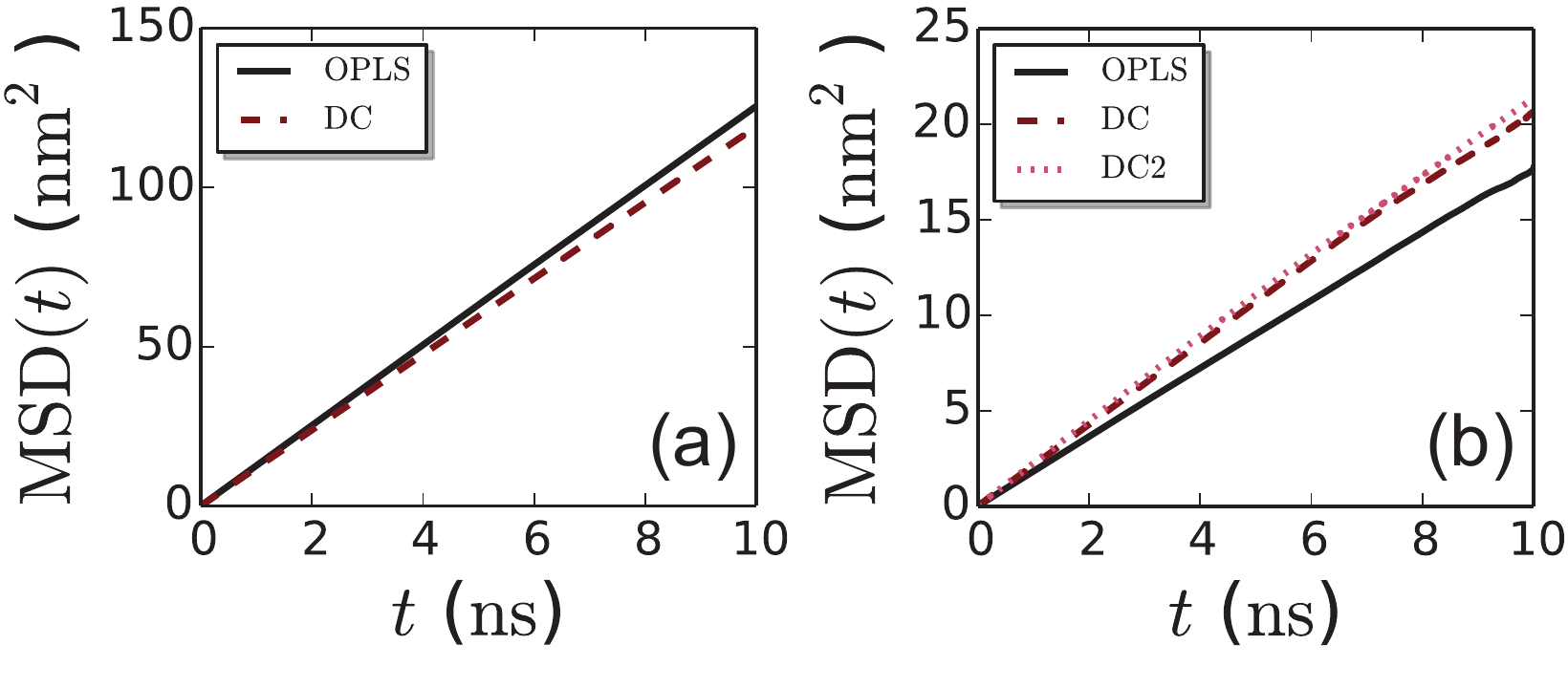}
\end{center}
\caption
{
Mean-squared displacement (MSD) as a function of time for the (a) methane and (b) ethane
models studied here. 
}
\label{fig:msd}
\end{figure}

%
To the extent that liquid structure determines dynamic properties in equilibrium,
the above results suggest that the DC models should yield liquid dynamics similar to the dipole-free OPLS models. 
To characterize the single-particle translational dynamics of each liquid, we compute the
mean-squared displacement (MSD) in each system.
The MSD is related to the diffusion coefficient, $D$, through the Einstein
relation, $6Dt=\lim_{t\rightarrow\infty}{\rm MSD}(t)$, such that similar MSDs
in two systems imply similar diffusion coefficients. 
The MSDs are shown in Fig.~\ref{fig:msd} for all systems under study. 
The dynamics of the DC models are slightly faster than the OPLS models,
which can be attributed in part to the slightly lower density of the DC models. 
The faster dynamics of the DC models is reflected in the diffusion coefficients, which we obtained by
linear fitting the long-time behavior of the MSD to $6Dt+c$.
This yields diffusion coefficients of
$D_{\rm OPLS}\approx 2.1\times 10^{-5}$~cm$^2$/s and
$D_{\rm DC}\approx  2.0\times 10^{-5}$~cm$^2$/s for the OPLS and DC
models of methane, respectively. 
Both models predict diffusion coefficients that are slightly smaller than that
obtained at $T=95.94$~K by Oosting and Trappeniers at coexistence~\cite{OT},
$D_{\rm exp}=3.01\times 10^{-5}$~cm$^2$/s.
The analogous diffusion coefficients for the ethane models are
$D_{\rm OPLS}\approx 0.30\times 10^{-5}$~cm$^2$/s and
$D_{\rm DC}\approx  0.35\times 10^{-5}$~cm$^2$/s, respectively. 
This further supports that the DC models diffuse slightly faster than the OPLS model,
and we also attribute this small difference to the slightly lower density of the DC system
at the same pressure. 
In this case, both models exhibit slightly slower diffusion than that determined
experimentally, $D_{\rm exp}\approx 0.8\times 10^{-5}$~cm$^2$/s,
by Gaven, Stockmayer, and Waugh at approximately 98~K~\cite{GSW}. 
%

\begin{figure}[tb]
\begin{center}
\includegraphics[width=0.48\textwidth]{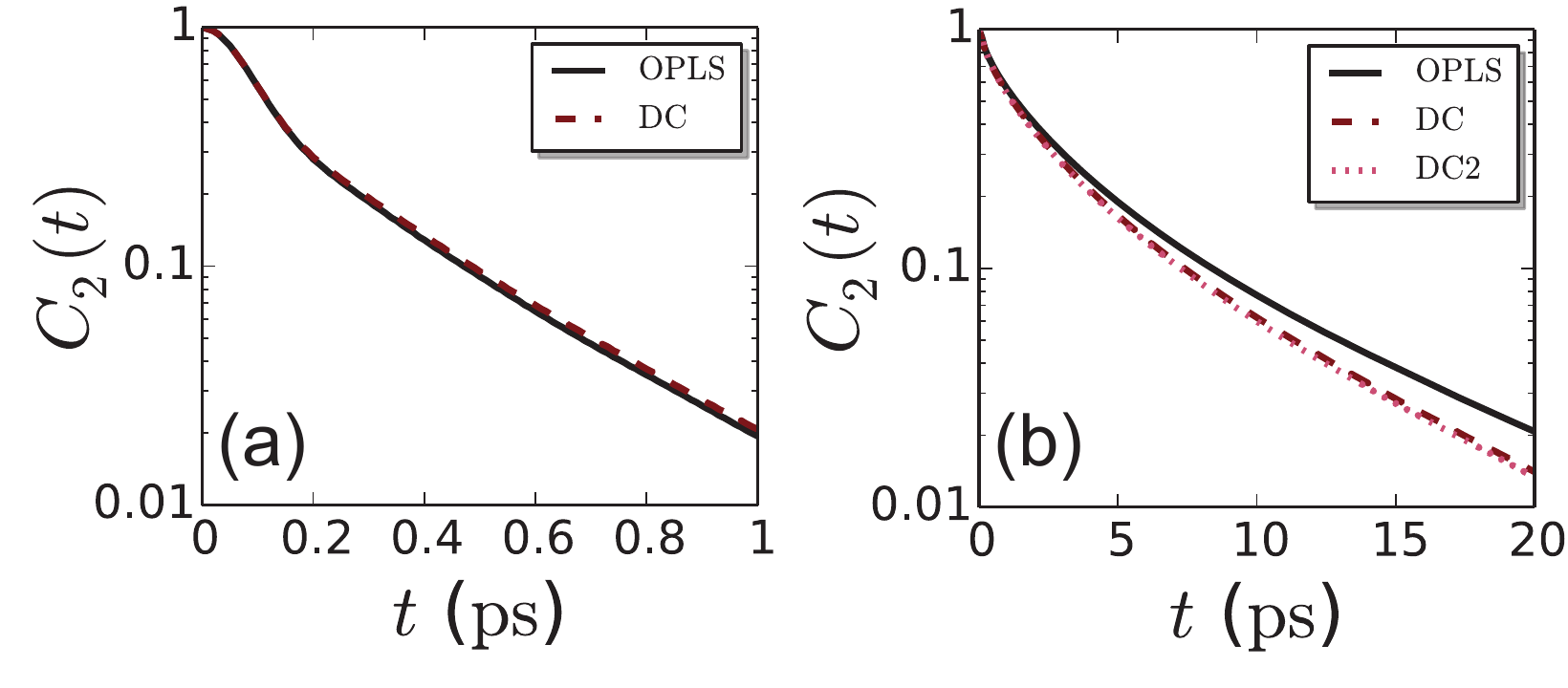}
\end{center}
\caption
{
Rotational time correlation functions, $C_2(t)$, for (a) methane and (b) ethane models
studied here. The methane $C_2(t)$ quantifies the rotation of the C-H bond vector,
while that for ethane quantifies the C-C bond rotation.  
}
\label{fig:rot}
\end{figure}

%
While the addition of a permanent dipole only slightly influences translational diffusion,
one might imagine that it could impact rotational motion. 
Therefore, we additionally examined single-molecule rotational dynamics by computing
the rotational correlation function
\begin{equation}
C_2(t) = \avg{P_2\para{ \mathbf{n}(t)\cdot\mathbf{n}(0)}},
\end{equation}
where $\mathbf{n}(t)$ is a C-H bond vector in the case of methane and the C-C bond vector in the case of ethane at time $t$
and $P_2(x)$ is the second order Legendre polynomial. 
These rotational correlation functions are shown in Fig.~\ref{fig:rot} for the methane and ethane models studied here. 
For methane, $C_2(t)$ is nearly identical for the OPLS and DC model, illustrating that the addition of a permanent dipole moment
does not significantly affect rotational motion in the liquid. 
Exponential fits to the long-time decay of $C_2(t)$ (0.2~ps to 2~ps) yield correlation times of $\tau_{\rm OPLS}\approx \tau_{\rm DC}\approx0.3$~ps,
further illustrating that the DC model minimally perturbs the dynamics of liquid methane. 
These correlation times are in good agreement with that of approximately 0.2~ps determined experimentally through Raman spectroscopy~\cite{gordon1965relations,McClung:JCP:1971}.
For ethane, $C_2(t)$ decays slightly faster in the DC and DC2 models than that for the OPLS model. 
The long-time decay of $C_2(t)$ for ethane (5~ps to 30~ps) is fit well with a bi-exponential, which we integrate to find the correlation time. 
This yields $\tau_{\rm OPLS}\approx3.48$~ps, $\tau_{\rm DC}\approx3.08$~ps, and $\tau_{\rm DC2}\approx3.03$~ps.
Performing the same fit on the experimental correlation function~\cite{wilde1981vibrational} yields a correlation time of 2.9~ps,
in good agreement with the DC model predictions. 
The addition of a permanent dipole moment in the DC models slightly speeds up the rotational dynamics of liquid ethane,
in addition to translational diffusion. 
While this can in part be attributed to a slightly lower density, the solvent's dynamic dielectric response, which involves rotational motion,
is inversely related to its dielectric constant, \ie \ higher dielectric constant liquids have faster dielectric response
when all other properties are the same~\cite{Zhao:JSP:2020}.
Thus, it may be expected that the DC models presented here will have slightly faster rotational dynamics
through their connection to dielectric relaxation. 
To summarize, the DC models yield a reasonable description of the
structure and dynamics of liquid methane and ethane, while also providing an accurate
representation of the static dielectric constant of each liquid.
%

\section{Density Fluctuations and Hard Sphere Solvation}
We now evaluate how altering the charge distribution of the methane and ethane models impact
solvation of small apolar solutes. 
To do so, we quantify density fluctuations in each liquid through the probability distribution, $P_v(N)$,
of observing $N$ heavy atoms in a spherical probe volume, $v$. 
For small $v$, $P_v(N)$ is expected to follow Gaussian statistics~\cite{LCW,HummerInfoTheory,pratt2002molecular,INDUS}. 
In this limit,
\begin{equation}
P_v(N) = \frac{1}{\sqrt{2\pi \avg{(\delta N)^2}_v}} \exp\brac{ - \frac{(N-\avg{N}_v)^2} {2\avg{(\delta N)^2}_v} },
\label{eq:gauss}
\end{equation}
where $\avg{N}_v=\rhob v$ is the average number of solvent molecules in $v$ at
a bulk density $\rhob$.
The variance in the number fluctuations, $\avg{(\delta N)^2}_v$, is given by
\begin{equation}
\avg{(\delta N)^2}_v = \int_v d\rb \int_v d\rb' \avg{\delta \rho(\rb) \delta\rho(\rb')},
\end{equation}
where the bulk density-density correlation function is
\begin{equation}
\avg{\delta \rho(\rb) \delta\rho (\rb')} = \rhob \omega_{\rm CC}(\len{\rb-\rb'})+\rhob^2\brac{g_{\rm CC}(\len{\rb-\rb'})-1}
\end{equation}
and $\omega_{\rm CC}(r)$ is the carbon-carbon intramolecular pair correlation function, equal to a delta function for methane~\cite{chandler1978structures,chandler1982new,chandler1988field}. 
Therefore, if $P_v(N)$ is Gaussian, we would expect the OPLS and DC models to yield
equivalent distributions, because both yield liquids with the same structure. 
%

\begin{figure}[tb]
\begin{center}
\includegraphics[width=0.48\textwidth]{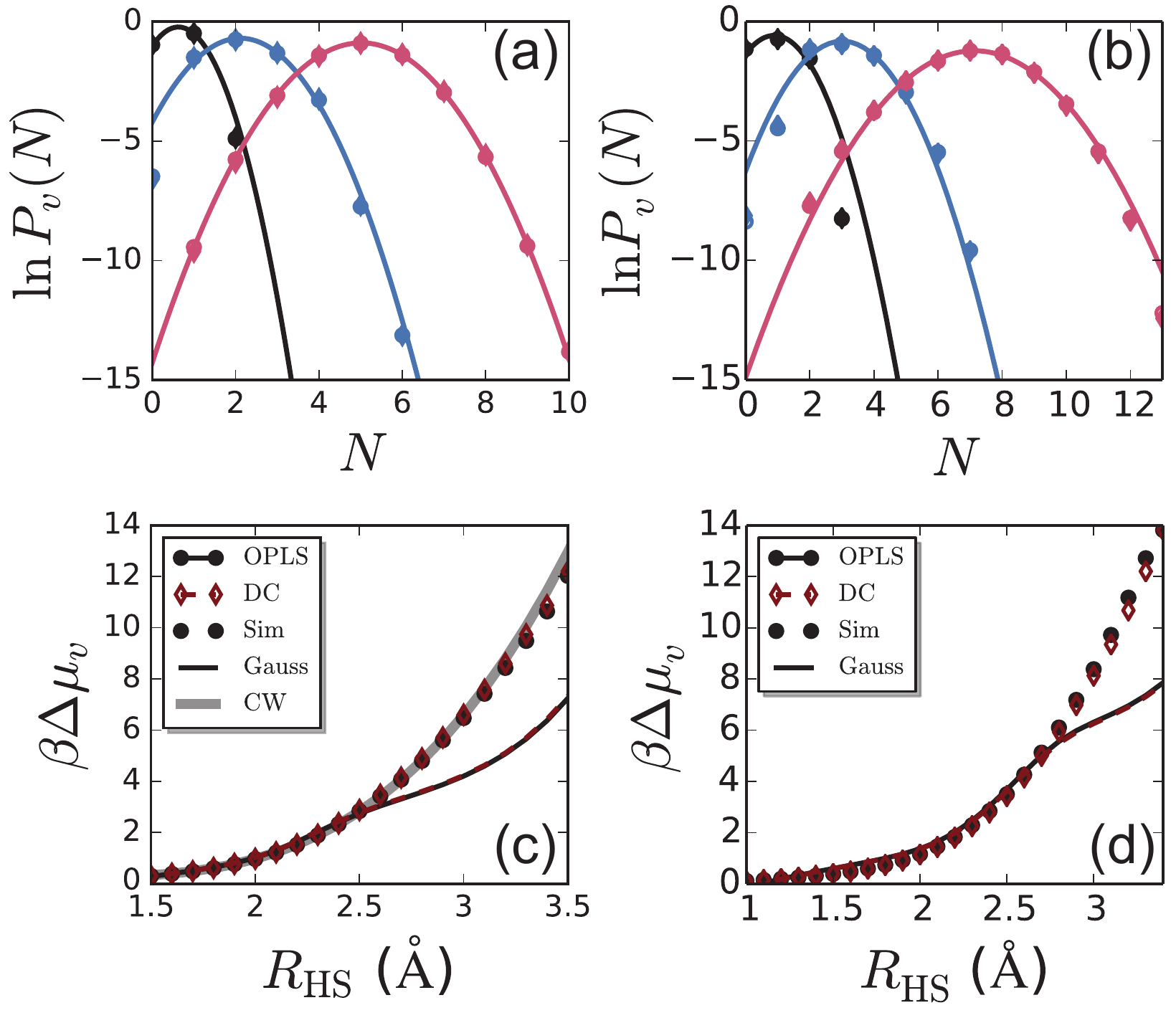}
\end{center}
\caption
{
(a,b) Probability distribution, $P_v(N)$, of the number of solvent molecules, $N$, within a spherical
volume, $v$, for (a) methane and (b) ethane models. From left to right, the spherical volumes
have radii of $R_{\rm HS}=2$~\AA, $R_{\rm HS}=3$~\AA, and $R_{\rm HS}=4$~\AA. 
OPLS model results are shown as circles, DC model results
are shown as diamonds. Solid lines correspond the predictions of Eq.~\ref{eq:gauss}. 
(c,d) Hard sphere solvation free energy, $\Delta\mu_v$, as of function of the solute radius for both models (points),
as well as their respective Gaussian approximations (thin solid/dashed lines). 
The thick gray line in (c) is the prediction of the theory of Chen and Weeks (CW)~\cite{Chen:JCP:2003}, Equation~\ref{eq:cw}.
}
\label{fig:pvn}
\end{figure}

%
The computed distributions, $P_v(N)$, are shown in Fig.~\ref{fig:pvn}a,b for liquid methane and ethane
models and representative spherical probe volumes, where $N$ corresponds to the number of carbon atoms in the probe volume. 
For small $v$, we find that the distributions are approximately Gaussian, and that the OPLS and DC models
yield equivalent distributions. 
This is expected based on the discussion above; both sets of models produce the same
$g_{\alpha\gamma}(r)$ and therefore the same density fluctuations. 
However, for larger volumes, close to $R_{\rm HS}\approx 3$~\AA \ and larger,
$P_v(0)$ is overestimated by the Gaussian prediction. 
The solvation free energy of a hard sphere of volume $v$, $\Delta \mu_v$,
can be obtained from the quantification
of density fluctuations using Widom's particle insertion~\cite{widom1963some}
\begin{align}
\beta\Delta \mu_v &= -\ln P_v(0) \\
&\approx \frac{\rhob^2 v^2}{2\avg{(\delta N)^2}_v} + \frac{1}{2}\ln\para{2\pi \avg{(\delta N)^2}_v},
\label{eq:gfe}
\end{align}
where the second line is obtained using the Gaussian approximation to $P_v(N)$ in Eq.~\ref{eq:gauss}.
Hard sphere solvation free energies as a function of solute size are shown in Fig.~\ref{fig:pvn}c,d
for liquid methane and ethane, along with the predictions of Eq.~\ref{eq:gfe}.
The free energies are in agreement for the two sets of charges, suggesting that the DC
models can be used for studying the solvation of apolar solutes. 
Moreover, the Gaussian approximation holds for hard sphere radii less than about 2.75~\AA,
suggesting that Eq.~\ref{eq:gauss} can be used to predict solvation free energies in this range of solute sizes. 
Above this size, the Gaussian approximation \emph{underestimates} the free energy,
as expected by the overestimate of $P_v(0)$ by the Gaussian approximation in Fig.~\ref{fig:pvn}a,b. 
These deviations from Gaussianity at low $N$ are also observed for hard sphere fluids~\cite{pratt2002molecular,pratt2001-HS,pratt2003-HS2}.
Within the perspective of Weeks-Chandler-Andersen (WCA) theory, the pair correlations in liquid methane and ethane are determined mainly
by the short-range, rapidly-varying repulsive cores of the molecular sites, while the slowly-varying, long-range attractions provide essentially
a uniform background potential~\cite{WCA,WidomScience,WCA-Science,chandler1978structures,chandler1982new}. 
Therefore, the molecular liquid can be accurately approximated by its purely short-ranged counterpart at the same bulk density. 
WCA also showed that the correlations within this short-ranged reference system can be further approximated by those
of an appropriately-chosen hard sphere reference solvent~\cite{Blip,WCA,WCA-Science}. 
Within this level of approximation, we can approximate the hard sphere solvation free energy, $\Delta \mu_v$,
by that in an appropriate hard sphere reference fluid. 
An analytic expression for this solvation free energy was derived by Chen and Weeks (CW)~\cite{Chen:JCP:2003},
\begin{align}
\beta\Delta \mu_v^{\rm CW} &= -\frac{\eta(2-7\eta+11\eta^2)}{2(1-\eta)^3} - \ln(1-\eta) \nonumber \\
&+\frac{18\eta^3}{(1-\eta)^3}\frac{R_{\rm HS}}{\sigma} - \frac{18\eta^2(1+\eta)}{(1-\eta)^3}\frac{R_{\rm HS}^2}{\sigma^2} \nonumber \\
&+ \frac{8\eta(1+\eta+\eta^2)}{(1-\eta)^3}\frac{R_{\rm HS}^3}{\sigma^3},
\label{eq:cw}
\end{align}
where $\eta=\pi\rhob\sigma^3/6$ is the packing fraction, $\sigma$ is the solvent hard core diameter,
and $R_{\rm HS}$ is the hard sphere solute radius.
Equation~\ref{eq:cw} was obtained following the `compressibility route'
described by CW, which was found to be the most accurate of several routes to the free energy explored in that work~\cite{Chen:JCP:2003}.
We set $\sigma=3.7$~\AA, which is roughly the hard sphere diameter of the
carbon atom plus half the C-H bond length
and is close to the first peak in $g_{\rm CC}(r)$. 
The predictions of Eq.~\ref{eq:cw} are shown as a gray solid line in Fig.~\ref{fig:pvn}c
and agree well with the simulation results for all values of $R_{\rm HS}$ studied here. 
For larger $R_{\rm HS}$ values, long-range solvent-solvent interactions become increasingly important,
but these can be accounted for using recent theoretical approaches~\cite{Remsing:2016ib}.
These results suggest that small-scale density fluctuations
in atomistic models of liquid methane are analogous to those of their hard sphere counterparts,
and solvation of small apolar solutes can be described within this level of approximation with reasonable accuracy. 
We expect that liquid ethane will follow similar principles --- apolar solvation can be described using
a hard diatomic fluid --- and we leave the extension of the CW theory~\cite{Chen:JCP:2003} and complementary approaches~\cite{DorDiatomics,chandler1978structures,pratt1980hydrophobic,chandler1982new}
to treat diatomic solvents with varying bond length for future work. 
%

\section{Free Energy of Hard Sphere Charging in Liquid Methane}
The results above demonstrate that the structure and dynamics of liquid methane and ethane,
and consequently apolar solvation in these two solvents, are essentially unaltered by introducing a small,
fixed dipole moment on each molecule. 
Thus, the DC models can describe the properties of liquid methane and ethane
as well as earlier dipole-free fixed charge models, with the additional advantage
of providing a reasonable description of the static dielectric constant. 
As an example of where dielectric response is significant and therefore differs between the two models,
we examine the process of charging hard sphere solutes in liquid methane.
We consider inserting a point charge at the center of a hard sphere of radius $R_{\rm HS}=3$~\AA \ in solution
and evaluate the corresponding free energies of charging the solute to a charge $Q$.
We obtain the charging free energy by linearly coupling the charge to a parameter $\lambda$,
such that $\lambda=0$ corresponds to the uncharged hard sphere and $\lambda=1$
indicates the fully charged solute. 
Through thermodynamic integration, the charging free energy is given by~\cite{Remsing:2016fq,Remsing:JSP:2019,Zhao:JSP:2020}
\begin{equation}
\Delta G^c(Q) = \int_0^1 d\lambda \int d\rb \int d\rb' \frac{\rho^Q(\rb) \rho^q_\lambda(\rb')}{\len{\rb-\rb'}},
\end{equation}
where
\begin{equation}
\rho^q_\lambda(\rb)=\avg{\rhoq(\rb;\Rbar)}_\lambda ,
\end{equation}
$\avg{\cdots}_\lambda$ indicates an ensemble average
over configurations sampled in solute charge state $\lambda Q$,
$\rhoq_\lambda(\rb;\Rbar)$ is the charge density in a single configuration $\Rbar$,
such that $\rhoq_\lambda(\rb)$ is the ensemble averaged solvent charge density at coupling parameter $\lambda$,
and $\rho^Q(\rb)=\rho^Q_{\lambda=1}(\rb)$
is the charge density of the solute in the fully coupled state ($\lambda=1$). 
For a point charge fixed at the origin, like those used here, $\rho^Q(\rb)=Q\delta(\rb)$,
which reduces the charging free energy to
\begin{equation}
\Delta G^c(Q) = Q \bar{v}^q(0),
\end{equation}
where
\begin{equation}
\bar{v}^q(\rb) = \int_0^1 d\lambda v^q_\lambda(\rb) = \int_0^1 d\lambda \int d\rb' \frac{\rho^q_\lambda(\rb')}{\len{\rb-\rb'}}
\end{equation}
is the $\lambda$-averaged electrostatic potential of the solvent. 
The charging free energies that we report are the ``Bulk'' free energies as defined
previously~\cite{Remsing:JPCL:2014,duignan2017electrostatic,Duignan:ChemSci:2017,Remsing:JSP:2019,Beck:2013,Doyle:2019aa},
\begin{equation}
\Delta G^c_{\rm Bulk}(Q) = \Delta G^c(Q) - Q\Phi^{\rm HW},
\end{equation}
where $\Phi^{\rm HW}$ is the electrostatic potential difference between the bulk liquid and vacuum (separated by a hard wall, for example),
which serves to appropriately reference the electrostatic potential to the vacuum.
Here, we approximate $\Phi^{\rm HW}$ by the potential difference across the liquid-vapor interface of each model, as done in previous work~\cite{Remsing:JPCL:2014,duignan2017electrostatic,Duignan:ChemSci:2017,Remsing:JSP:2019,Ashbaugh:2000,Palmeri:2013,Beck:2013}.
For the models studied here, this is also equal to the Bethe potential of the model because
there is no preferential orientation of dipole moments at the liquid-vapor interface~\cite{Harder:2008,Wilson:1989,Kathmann:2011,Remsing:JPCL:2014,duignan2017electrostatic,Duignan:ChemSci:2017,Remsing:JSP:2019,Doyle:2019aa}.
We compare the simulation results to the Born model of charging~\cite{Born},
\begin{equation}
\Delta G_{\rm Born}(Q) = -\frac{Q^2}{2 R_{\rm B}}\para{1-\frac{1}{\varepsilon}},
\label{eq:born}
\end{equation}
where $Q$ and $R_{\rm B}$ are the charge and Born radius of the ion. 
While the Born radius can be estimated from simulations in several ways~\cite{Remsing:JPCL:2014,duignan2017electrostatic,Duignan:ChemSci:2017,Remsing:JSP:2019},
we approximate it by the hard sphere radius of the solute, $R_{\rm B}\approx R_{\rm HS}=3$~\AA. 
%

\begin{figure}[tb]
\begin{center}
\includegraphics[width=0.43\textwidth]{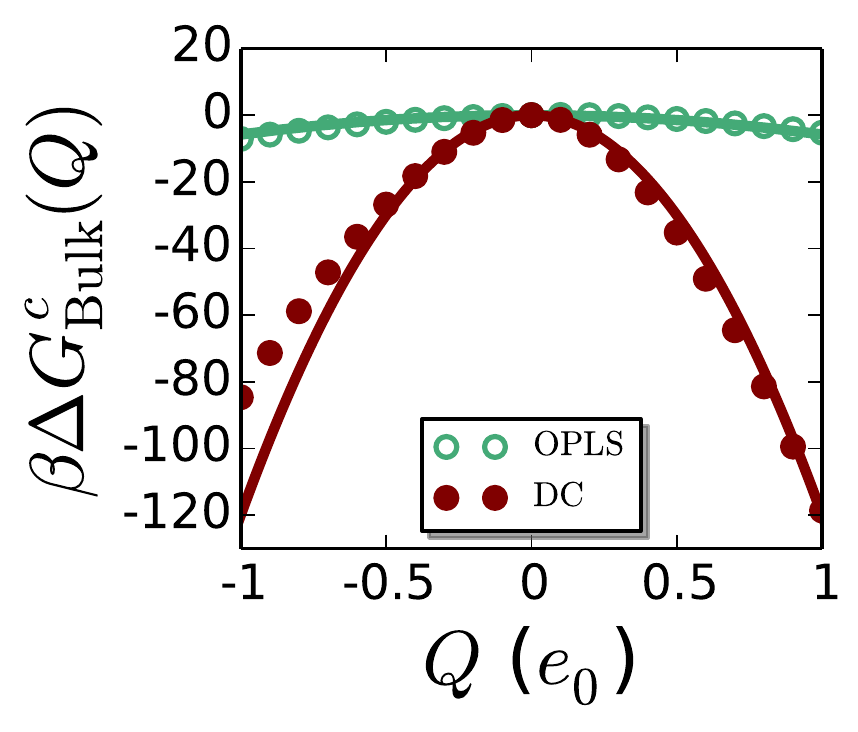}
\end{center}
\caption
{
Charging free energy as a function of the solute charge.
Solid lines are predictions of the Born model with $R_{\rm B}=3$~\AA.
The Born model curve for the OPLS methane model uses a larger dielectric constant (1.02)
than that explicitly calculated for the uniform bulk liquid. 
}
\label{fig:cfe}
\end{figure}

\begin{figure}[tb]
\begin{center}
\includegraphics[width=0.48\textwidth]{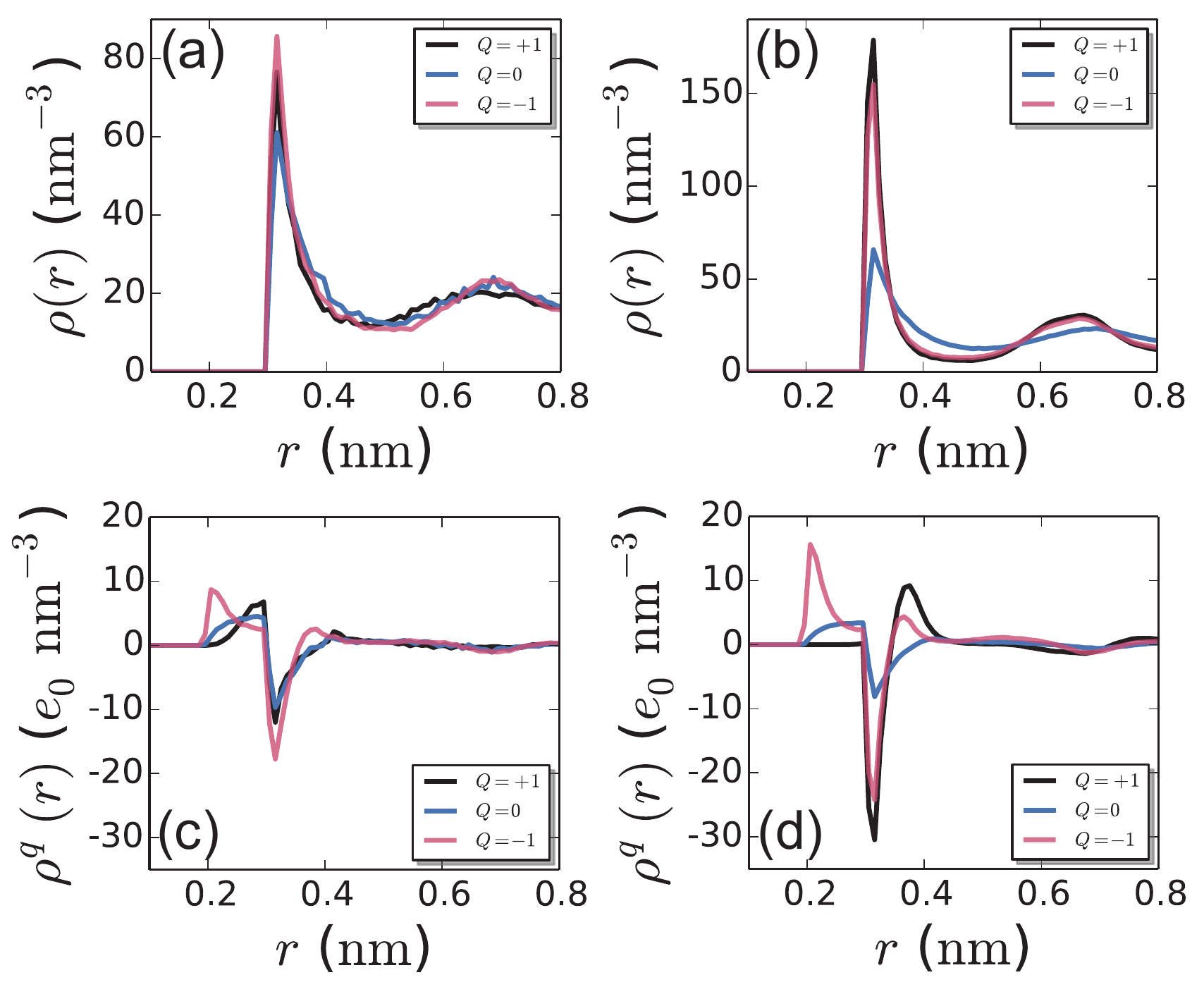}
\end{center}
\caption
{
Nonuniform (carbon) density profiles, $\rho(r)$, for the (a) OPLS and (b) DC methane models
around a hard sphere with radius $R_{\rm HS}=3$~\AA \ and charges of $Q=0,\pm1$,
as well as the corresponding charge densities, $\rhoq(r)$, for the (c) OPLS and (d) DC models. 
}
\label{fig:dens}
\end{figure}

%
The charging free energies are shown in Fig.~\ref{fig:cfe} for both models. 
The Born model provides a good approximation to the magnitude of the charging free energies, 
although the simulated free energies display a slight asymmetry with respect to $Q$. 
This asymmetry is becoming increasingly well understood and arises
from the asymmetric charge distribution of the molecular model~\cite{Remsing:JPCL:2014,duignan2017electrostatic,Duignan:ChemSci:2017,Remsing:JSP:2019,Mukhopadhyay:2012},
in addition to the asymmetric nature of the solute-solvent excluded volume interactions~\cite{Shi:2013,D0CP04148C}. 
Importantly, $\Delta G^c_{\rm Bulk}(Q)$ obtained for the DC model is roughly a factor of 20 larger in magnitude (more favorable)
than that obtained for the OPLS model. 
This is consistent with the inability of the dipole-free OPLS model to describe the dielectric response of the solvent to
charged and polar solutes. 
A similar increase in the charging free energy magnitude from the dipole-free OPLS model to the DC model can be expected for dipolar
solutes as well, based on the Bell model~\cite{bell1931electrostatic,Zhao:JSP:2020}, the analogue of the Born model for dipolar hard sphere solvation in a dielectric.

The inability of the dipole-free models to respond to solute charging
is further demonstrated by the densities and charge densities in Fig.~\ref{fig:dens}. 
The density, $\rho(r)$, of the OPLS methane molecules (Fig.~\ref{fig:dens}a) displays only slight changes upon charging the solute,
while $\rho(r)$ for the DC model (Fig.~\ref{fig:dens}b) displays a large response to charging; the first peak triples in magnitude, for example. 
The differences in the nonuniform density ultimately arise from the ability of the DC model to interact with charged solutes via
charge-dipole interactions, while these are absent in the OPLS model. 
This point is further exemplified by the charge densities for the OPLS and DC models shown in Fig.~\ref{fig:dens}c and~\ref{fig:dens}d, respectively.
The OPLS model does not have a permanent dipole to preferentially orient, so little change is observed in the solvent structure
 as the charge state of the solute is varied.
The DC models exhibit very significant differences in the charge densities around the cationic, anionic, and uncharged hard spheres,
which originate from the preferential orientation of the solvent dipole moments in the solvation shell in response to a solute charge.
A large positive peak is observed close to the anionic solute, and this peak is replaced by a large negative peak
around the cationic solute, as may be expected for dipolar molecules with opposite orientations in the solvation shell. 
This suggests that the DC models developed here can provide an approximate microscopic description of dielectric response that is lacking
in conventional hydrocarbon models. 
%

\section{Solvation Free Energies of Idealized Ionic Solutes in Liquid Methane}
\begin{figure}[tb]
\begin{center}
\includegraphics[width=0.48\textwidth]{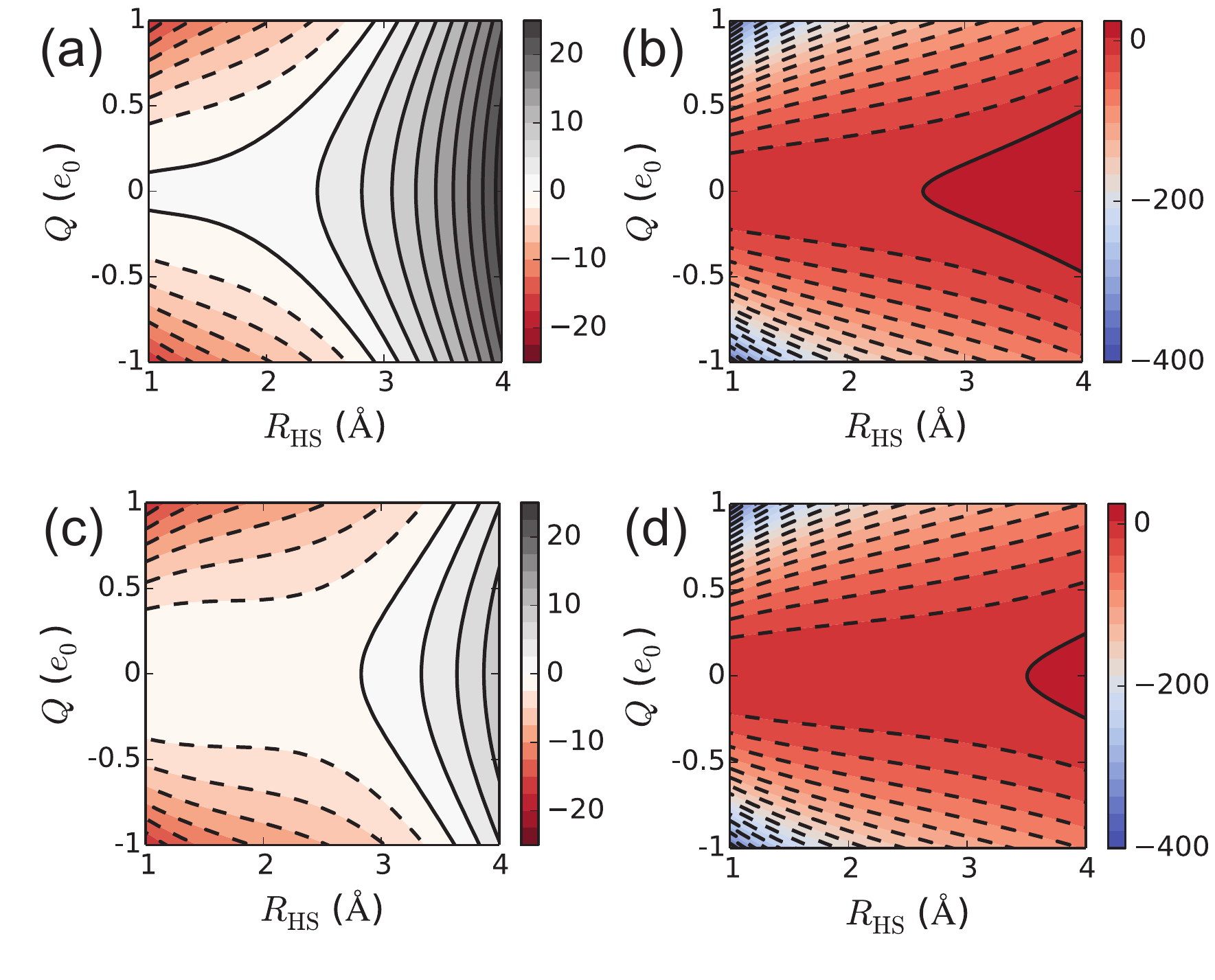}
\end{center}
\caption
{
(a,b) Total solvation free energies for charged hard spheres, $\beta\Delta G(R_{\rm HS},Q)$, in the (a) dipole-free OPLS
and (b) DC models of methane, predicted using Eq.~\ref{eq:chs}
(c,d) The total solvation free energies for charged hard spheres with a Lennard-Jones attractive potential for
(c) OPLS and (d) DC models of liquid methane. 
The contribution to the free energy from turning on the Lennard-Jones solute-solvent attraction
is estimated following Eq.~\ref{eq:step}.
Solid/dashed contour lines indicate positive/negative free energies. 
}
\label{fig:total}
\end{figure}

The results of the previous two sections can be combined to estimate the solvation free energy
of charged hard spheres in liquid methane using the OPLS and DC models. 
The total solvation free energy of a charged hard sphere, $\Delta G(R_{\rm HS},Q)$,
can be approximated by a combination of the CW and Born theories for inserting the solute core and
subsequently charging it, respectively,
\begin{equation}
\Delta G(R_{\rm HS},Q) \approx \Delta\mu_v^{\rm CW}(R_{\rm HS}) + \Delta G_{\rm Born}(R_{\rm HS},Q),
\label{eq:chs}
\end{equation}
where we have emphasized that the first term does not depend on solute charge 
and we use the hard sphere radius as the Born radius.
The total solvation free energies are shown in Fig.~\ref{fig:total} for the two methane models. 
For the dipole-free model, only small charged hard spheres have a favorable solvation free energy.
In contrast, the DC model favorably solvates monovalent ions of all sizes studied here,
as well as partially charged ions approaching $\len{Q}=e_0/2$ for $R_{\rm HS}<3$~\AA.
%

%
We can also add attractive van der Waals-like interactions between the solute and solvent
in order to better mimic a physical solute. 
This is accomplished by considering one additional step at the end of the solvation process,
in which the solute-solvent attractive interaction $u_1(r)$ is turned on, after charging. 
Within linear response theory, the free energy change of turning on this attractive interaction is
\begin{equation}
\Delta G_1 \approx \int d\rb \rho(r) u_1(r),
\label{eq:lrt}
\end{equation}
where $\rho(r)$ is the solvent density around the solute and the attractive interaction is
given by the attractive portion of a Lennard-Jones potential,
\begin{eqnarray}
u_1(r)= \Bigg\{ \begin{array}{ll}
-\epsilon, \ \ \ \ \ \ \ \ \ \ \ \ \ \ \ \ \ \ \ \ \ \ \ \ \ \ \  r < R_{\rm HS} \\
4\epsilon \brac{ \para{\frac{R_{\rm HS}}{r}}^{12} - \para{\frac{R_{\rm HS}}{r}}^{6} }, \ r\ge R_{\rm HS}\\  \end{array}
\nonumber
\end{eqnarray}
In order to examine the qualitative effects of adding $u_1(r)$ for many values of $R_{\rm HS}$,
we further approximate $\Delta G_1$ by taking a sharp-kink approximation
to the induced solvent density, $\rho(r)=\rhob\Theta(r-R_{\rm HS})$,
where $\Theta(r)$ is the Heaviside function.
This enables the free energy to be evaluated analytically,
\begin{equation}
\Delta G_1 \approx -\frac{32}{9}\pi\beta\tilde{\epsilon}\rhob R_{\rm HS}^3.
\label{eq:step}
\end{equation}
The effective well-depth, $\tilde{\epsilon}=0.86$~kJ/mol, was chosen so that $\Delta G_1$ obtained via Eq.~\ref{eq:step}
agrees with that determined by evaluating Eq.~\ref{eq:lrt} using the simulated density of the OPLS model for $R_{\rm HS}=3$~\AA \ 
and $\epsilon=0.7$~kJ/mol.
We show the total solvation free energy of charged hard spheres with LJ attractions in Fig.~\ref{fig:total}c,d
within this crude level of approximation for the two liquid methane models. 
As may be expected, attractive interactions ensure that small uncharged solutes are favorably solvated.
Large, even partially charged attractive hard spheres are unfavorably solvated in the OPLS model.
Attractive solutes in the DC model are favorably solvated for nearly the entire range of $R_{\rm HS}$ studied here. 
To summarize, the predicted $\Delta G(R_{\rm HS},Q)$ highlight the importance of dielectric effects in determining
even the qualitative behavior of the thermodynamics governing simple solute solvation in liquid hydrocarbons,
in addition to the large quantitative differences between the two types of models. 
%

\section{Solvation Free Energy of Water in Liquid Methane}
{The importance of dielectric response in solvation thermodynamics is also apparent
when the solute is neutral but polar, as is the case for water. 
Water is prevalent on Titan, as a subsurface ocean as well as ice on its surface~\cite{TitanChapter},
and understanding the solvation thermodynamics of water in liquid methane is a prerequisite for
predicting water's role in more complex chemical processes. 
In this section, we compute the solvation free energy of a water molecule in the OPLS and DC
models of liquid methane. }
{We model water using the SPC/E model~\cite{SPCE}, which interacts with methane via
a LJ potential centered on the oxygen site
and point charges $q_{\rm H}$ and $q_{\rm O}=-2q_{\rm H}$ located on
the hydrogen and oxygen sites, respectively.
We divide the solvation free energy into two components,
the free energy of inserting the LJ core into the solvent, $\Delta G^{\rm LJ}$, and
the free energy of charging the H and O sites, $\Delta G^\mu$, where $\mu$
indicates the dipole moment of the solute. 
The total solvation free energy is then given by $\Delta G = \Delta G^{\rm LJ} + \Delta G^\mu$.}
{The first, LJ term in the solvation free energy can be obtained by standard Widom particle insertion~\cite{widom1963some},
\begin{equation}
\beta\Delta G^{\rm LJ} = -\ln \avg{e^{-\beta \Delta U_{\rm LJ}}}_{\rm B},
\label{eq:widom}
\end{equation}
where the subscript ${\rm B}$ indicates that the average is performed over configurations of the bulk
solvent,
\begin{equation}
\Delta U_{\rm LJ} = \sum_{i=1}^N u_{i{\rm W}}(\len{\rb_i - \rb_{\rm W}})
\end{equation}
is the interaction energy between the LJ core and the $N$ atoms of the solvent, and
$u_{iW}(r)$ is the LJ interaction potential between solvent atom $i$ and LJ core (W),
each located at $\rb_i$ and $\rb_{\rm W}$, respectively. }
{The insertion of water's LJ core into liquid methane does not involve water-methane electrostatic interactions.
We can then expect that OPLS and DC models will yield similar values of $\Delta G^{\rm LJ}$,
as is the case for hard sphere insertion. 
Indeed, evaluating Eq.~\ref{eq:widom} over 50~ns trajectories in each methane model results in similar
LJ core insertion free energies, as listed in Table~\ref{tab:water}.}
{To determine $\Delta G^\mu$,
we first define the electrostatic interaction energy between the water solute and liquid methane as
\begin{equation}
\Delta U_\mu = \sum_{i=1}^N \sum_{j=1}^3 \frac{q_i q_j}{\len{\rb_i - \rb_j}},
\end{equation}
where $q_i$ is the charge on solvent atom $i$ and $q_j$ is the charge on solute atom $j$,
located at $\rb_i$ and $\rb_j$, respectively. 
We then linearly couple this energy to a coupling parameter, $\lambda$, such that $\lambda=0$
corresponds to the uncharged LJ core and $\lambda=1$ corresponds to a fully-charged SPC/E
water molecule. 
Through thermodynamic integration, the free energy of charging the water molecule is given by
\begin{equation}
\Delta G^\mu_{\rm sim} = \int_0^1 d\lambda \avg{\Delta U_\mu}_\lambda,
\end{equation}
where $\avg{\cdots}_\lambda$ indicates an ensemble average over configurations in the system
with coupling parameter $\lambda$ and the subscript ${\rm sim}$ indicates that the free energy
is determined in a finite-size simulation box. 
To correct for the finite size of the simulation cell, we add a long range correction to the free energy
determined by the Bell model, which describes solvation of a fixed dipole inside a cavity of radius $R$ within a dielectric,
\begin{equation}
\Delta G^\mu_{\rm Bell}(R) = -\frac{\varepsilon-1}{2\varepsilon+1} \frac{\mu^2}{R^3},
\end{equation}
where $\mu$ is the dipole moment of the solute. 
The finite size correction is obtained by setting the radius equal to the average length of the cubic simulation cell, $R=L$.
Then, $\Delta G^\mu_{\rm Bell}(L)$ accounts
for solute-solvent interactions beyond the box length,
such that $\Delta G^\mu = \Delta G^\mu_{\rm sim} + \Delta G^\mu_{\rm Bell}(L)$.}
%

\begin{figure}[tb]
\begin{center}
\includegraphics[width=0.48\textwidth]{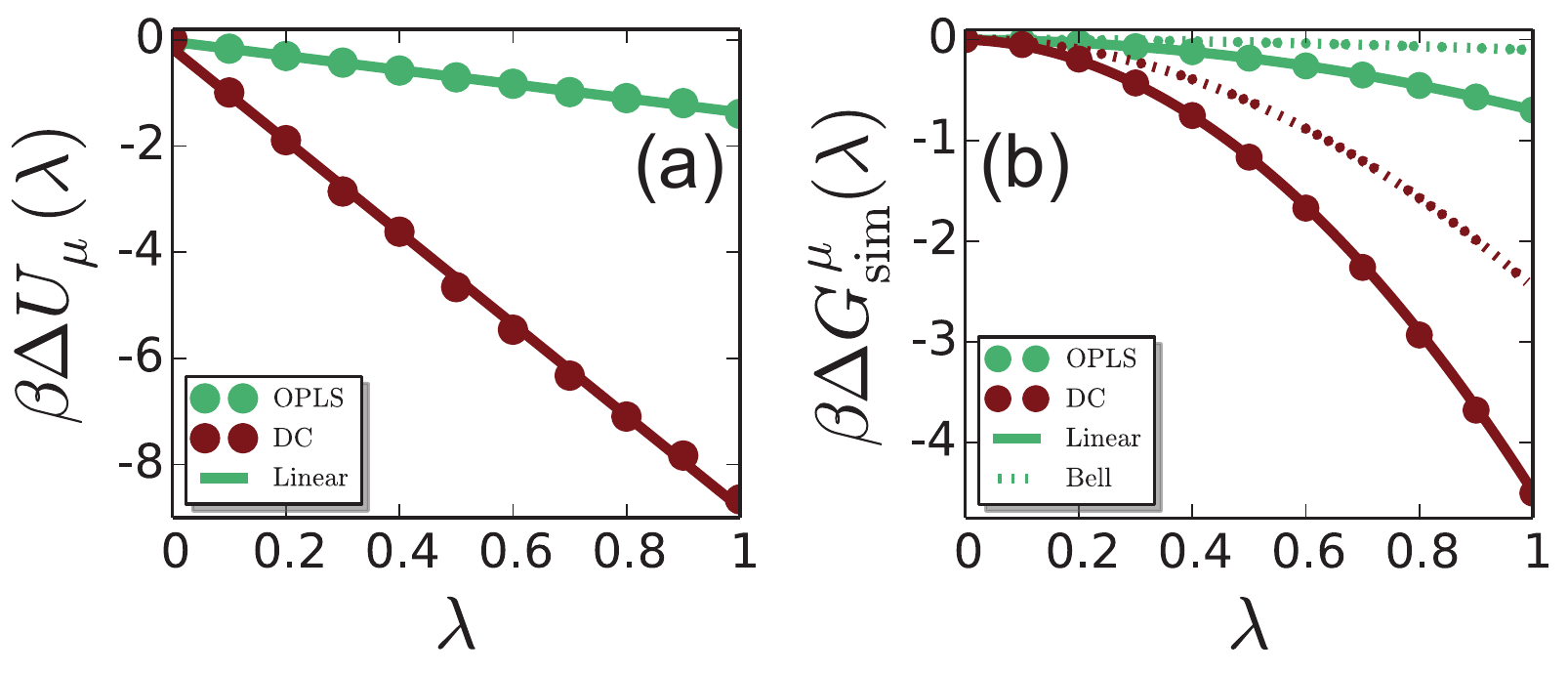}
\end{center}
\caption
{
{(a) Average water-methane electrostatic interaction energy evaluated as a function of $\lambda$
and 
(b) the corresponding charging free energy evaluated by thermodynamic integration.
Data points correspond to simulation results,
solid lines correspond to linear fits in (a) and their integral in (b),
and the dashed line in (b) corresponds to predictions of the Bell model for
a solute radius of $R=2.5$~\AA \ and a box size equal to that of the simulation cell,
$\Delta G^\mu_{\rm Bell,sim} = \Delta G^\mu_{\rm Bell}(R)-\Delta G^\mu_{\rm Bell}(L)$.}
}
\label{fig:TI}
\end{figure}

%
{The $\lambda$-dependent average interaction energy, $\avg{\Delta U_\mu}_\lambda$,
is linear to a good approximation in both the OPLS and DC models (Fig.~\ref{fig:TI}a).
This suggests that linear response theories are applicable to the solvation of small polar molecules
in liquid methane, and the free energy is quadratic in $\mu$, as shown in Fig.~\ref{fig:TI}b
and predicted by dielectric continuum theories.
The response of the DC model is larger than that of the OPLS model, consistent
with the DC model's larger dielectric constant. 
The finite size-corrected free energy is roughly
eight times larger in magnitude in the DC model, Table~\ref{tab:water}.
This is again consistent with the larger dielectric response of the DC model and
expectations from dielectric continuum theory. 
However, the Bell model itself underestimates the charging free energy in both the OPLS and DC models,
where the Bell model results were obtained using $R=2.5$~\AA \ for both models, and a higher dielectric constant of 1.02 for the OPLS model,
as done above for the Born model. 
The differences can be attributed to the neglect of short range molecular details and higher order multipolar interactions in the Bell model.}
%

\begin{table*}
\caption{
\label{tab:water}
{Solvation free energies and its components for a SPC/E water molecule in liquid methane. 
{Error bars represent one standard error.}}
}
\vspace{0.6cm}
\centering
\begin{tabular*}{\hsize}{@{\extracolsep{\fill}}lccccc}
\hline
Model & $\beta\Delta G^{\rm LJ}$ & $\beta\Delta G^{\rm \mu}_{\rm sim}$ & $\beta\Delta G^{\rm \mu}$ & $\beta\Delta G$ & $\beta\Delta G_{\rm Bell}$  \\
\hline
OPLS & $-2.387\pm0.007$ & $-0.70\pm0.01$ & $-0.77\pm0.01$ & $-3.16\pm0.01$ & $-0.18$  \\[1.0ex]
DC & $-2.358\pm0.005$ & $-4.50\pm0.02$ & $-6.26\pm0.02$ &  $-8.75\pm0.02$ & $-4.25$    
\end{tabular*}
\end{table*}

%
{The total solvation free energies, $\Delta G$, obtained by summing the LJ and charging free energies are
listed in Table~\ref{tab:water}.
The resulting solvation free energy in the OPLS model is nearly three times smaller in magnitude than that of the DC model.
This illustrates that the lack of a physical dielectric response in symmetric models like OPLS can
result in predictions that significantly underestimate solvation free energies and, consequently, solution-phase binding free energies.
The significant difference in solvation free energies found here
emphasizes the importance of including dielectric response in models of liquid hydrocarbon solutions.}
%

\section{Conclusions}
We have developed models of liquid methane and ethane in which molecular charge symmetry
is broken by creating a fixed dipole moment in order to describe the dielectric constant of the liquid. 
The resulting DC models accurately describe the structure and dynamics of the liquids,
while gaining the ability to estimate dielectric response, in a mean-field-like manner, by replacing
the polarizability fluctuations of the real system with effective permanent dipole moments.
{Finally, we demonstrated that these new models can describe solvation of charged and polar solutes,
such as water, for which solvent dielectric response is critical.}
We expect these new DC models to be useful in the study of solvation and assembly of polar and charged solutes
in liquid methane and ethane, both of which require a description of dielectric response~\cite{bader1992computer,hirata1983interionic,AcetonitrilePMF,Remsing:2016ib,gao2020short}.
In particular, there is great interest in understanding chemistry that could be occurring
in the liquid hydrocarbon lakes on the surface of Titan. 
The first step in achieving this goal is understanding the solvation structure and thermodynamics of
relevant molecules. 
Such information is difficult to gather experimentally, due to the cryogenic conditions needed to mimic Titan's lakes,
and predictive molecular simulations enabled by DC models
will play an important role in characterizing solvation and assembly in these liquid hydrocarbon environments.

\section*{Acknowledgements}
We thank Pratip Chakraborty and Ryan Nival for helpful discussions. 
Input files necessary to reproduce the simulations in this work are available
at github.com/remsing-group/MethaneEthaneDCModels.
This work is supported by the
National Aeronautics and Space Administration under grant number 80NSSC20K0609 issued through the NASA Exobiology Program.
We acknowledge the Office of Advanced Research Computing (OARC) at Rutgers,
The State University of New Jersey
for providing access to the Caliburn cluster
and associated research computing resources that have contributed to the results reported here.
This article is dedicated to Mike Klein on the occasion of his 80th birthday. 
\bibliographystyle{ieeetr}

\end{document}